\documentclass[12pt,preprint,dvips,deluxetable]{aastex}
\usepackage{amsmath}
\usepackage{xcolor}
\usepackage[multiple]{footmisc}
\def \ra{R.A.}
\def \dec{Dec.}

\shorttitle{The WIRO Wide-Field Camera and Remote Operations}
\shortauthors{Findlay, J.~R., Kobulnicky, H.~A., et al.}

\begin{document}
\slugcomment{2016 April 8}

\title{A Wide-Field Camera and Fully Remote 
Operations at the Wyoming Infrared Observatory}

\author{
Joseph~R.~Findlay\altaffilmark{1},
Henry~A.~Kobulnicky\altaffilmark{1}
James~S.~Weger\altaffilmark{1},
Gerald~A.~Bucher\altaffilmark{1},
Marvin~C.~Perry\altaffilmark{1},
Adam~D.~Myers\altaffilmark{1},
Michael~J.~Pierce\altaffilmark{1},
Conrad~Vogel\altaffilmark{2}
}

\altaffiltext{1}{Department of Physics \& Astronomy, University 
of Wyoming, Laramie, WY 82070, USA}
\altaffiltext{2}{Steward Observatory, University of Arizona, 
Tucson, AZ 85721, USA }
\email{joseph.findlay@uwyo.edu}

\begin{abstract} Upgrades at the 2.3 meter Wyoming Infrared
Observatory telescope have provided the capability for
fully-remote operations by a single operator from the
University of Wyoming campus. A line-of-sight 300 Megabit 
$\mathrm{s}^{-1}$ 11 GHz radio link provides high-speed internet for
data transfer and remote operations that include several
real-time video feeds. Uninterruptable power is ensured by a
10 kVA battery supply for critical systems and a 55 kW
autostart diesel generator capable of running the entire
observatory for up to a week.  Construction of a new
four-element  prime-focus corrector with fused-silica
elements  allows  imaging over a 40\arcmin\
field-of-view with a new $4096^2$ UV-sensitive prime-focus
camera and filter wheel.  A new telescope control system
facilitates the remote operations model and provides
20\arcsec\ rms pointing  over the usable sky. Taken
together, these improvements pave the way for a  new
generation of sky surveys supporting space-based missions and
flexible-cadence observations advancing emerging
astrophysical priorities such as planet detection, quasar
variability, and long-term time-domain campaigns.  
\end{abstract}


\section{Introduction}

The Wyoming Infrared Observatory (WIRO) 2.3-meter classical
Cassegrain telescope was constructed by the University of
Wyoming  in 1977 as (briefly) the world's largest
infrared-optimized telescope and the first fully
computer-controlled telescope \citep{GH78}.  Situated at
9656 feet (2935 m) above sea level atop Mt.~Jelm, 25 miles
southwest of Laramie, WY, it is one of the highest research
observatories in the continental U.S. WIRO  remains one of
the largest telescopes that is 100\% operated by a single
institution.  The equatorial yoke-mounted telescope has an
f/27 Cassegrain focus (with optional chopping secondary) 
and an f/2.1 prime focus, providing plate scales of
3\farcs32 mm$^{-1}$ and 43\arcsec\ mm$^{-1}$, respectively. 
Among its achievements, it performed early infrared
observations of dust in evolved stars, \citep{Hackwell79, 
Gehrz80} and some of the first maps of infrared sources
\citep{Gehrz82, Hackwell82}. More recently it has been used
to map the star formation rates in the local universe
\citep{Dale10} and complete a statistical census of orbital
parameters in a large sample of massive binary stars
\citep{K14}.  WIRO has enabled hands-on telescope and
instrumentation training for  generations of graduate and
undergraduate students while contributing to over 45
doctoral dissertations and over 115 refereed journal
articles.\footnote{The WIRO website at
\url{http://physics.uwyo.edu/\textasciitilde WIRO} maintains a list
publications. }  

A series of recent upgrades to the telescope and instrument
suite have equipped WIRO for a new generation of 
large-scale surveys and synoptic programs.  Section 2
introduces the new control system installed in November of 2008,
replacing the 1980's-era telescope electronics and control loop
software.  Section 3 describes a new prime-focus corrector
built and implemented in 2004.  Section 4 details a new
prime-focus assembly and CCD camera implemented in 2014. 
Section 5 highlights  infrastructure allowing fully-remote
single-operator  observations from the University of Wyoming
campus.   Together, these improvements have allowed more
efficient and effective use of  this small- mid-sized
telescope.  

\section{Telescope Control System}

In 2008 General Dynamics (GD) SATCOM, Inc.\ designed and
built a control system, in consultation with WIRO staff, for
the 2.3 meter telescope and dome.  It was installed in
November 2008 by GD SATCOM and WIRO staff as the most
significant upgrade of the telescope infrastructure since
its construction.   Encoders on the \ra\ and \dec\ axes are
located on worm gears that drive the bull gears on each axis
at a rate of 0.5\degr\ per rotation of the worm gear.   The
old encoders (14-bits per turn)  were replaced with 37-bit
(25-bits per turn plus 12 bits to encode absolute position)
Heidenhain ROQ437 multiturn encoders.  
The original drive motors and pre-load
motors\footnote{Drive motors on each axis are Inland Torque
model T-7202-E with Inland Tachometer Generator model
TD-5110-B tachometers for rate control loop feedback. 
Tachometers are mounted on the same shaft as the torque
motors.  Pre-load motors are Inland Torque Motor model
T-7203-C. Inland Torque is now part of Kollmorgen and these 
parts are no longer in production. The interested reader should
contact Kollmorgen for advice on equivalent products.}
were retained.  A Clifton Precision synchro-brushless 
resolver model 11-BHW\footnote{Clifton Precsision
is now part of MOOG.} 
mounted on the rotating dome ring reports dome azimuth to 0.1 degrees.

Amplifiers for telescope and dome motors reside in a custom
chassis, provided by GD SATCOM, located in the WIRO control
room.  The chassis contains pulse width modulation servo
motor  amplifiers, an Advanced Motion Controls model 
25A8\footnote{This product is no longer available
to new applications, contact Advanced Motion 
Controls to identify an equivalent.}
for low-level velocity loop control of \ra\ and \dec, and a
central control unit (CCU) for position loop control of
each axis. Communications between the CCU and
telescope control computer, located in an adjacent rack and
running real time linux (QNX), occurs over a short RS422
connection.  A six-parameter pointing model within the GD
SATCOM control loop applies differential corrections for
small misalignment of each of the telescope axes in each of
two dimensions, including a  term for zero points of each
axis.  Dome azimuth is synchronized to telescope location
under normal sidereal tracking modes, with a deadband of
three degrees to inhibit constant dome motion while
preventing occultation.   

Operators control the telescope from a linux  PC serving as
the front end to the GD SATCOM telescope control computer. 
The text-driven front-end C code  is derived from the
previous WIRO telescope control system developed at UW \citep{Spillar93},
which, in turn, has roots in the 1976-era FORTH  control
code originally implemented at WIRO.  The front-end C code
passes telescope and dome commands to the telescope control
computer and reads telescope status from the telescope
control computer using formatted text strings over
ethernet.  The front-end interface allows arbitrary \ra\ \&
\dec\ offsets, which are stored upon shutdown, and implements a
capability for non-sidereal tracking inherent in the SATCOM
control loop; users may specify  non-sidereal rates in \ra\
and \dec\ in arc-seconds per second.   

Telescope pointing is $\approx$20\arcsec\ RMS over the sky
at zenith angles less than 60\degr.       Open-loop
exposures of up to five minutes are routinely obtained with
negligible tracking error. Maximum slew rates are 40 degrees
per minute on each axis. Settle times are about eight
seconds.  The encoders and control system electronics  have
proven robust against lightning during summer storms. An
extensive system of grounding and surge suppression (Section
5) has also helped to minimize electrical damage.

\section{A New Wide-Wield Prime-Focus Corrector}

The 2.3-meter primary mirror of WIRO is an f/2.1 paraboloid
and, as such, it  suffers from off-axis coma that severely
limits its usable field of view.  In the late 1980's Ed Loh
and Earl Spillar designed a four-element prime  focus
corrector for WIRO based upon the designs of \citet{Wynne65,
Wynne73} that provided a usable field of about 
8\arcmin.  Since that time, much larger format CCD
detectors have become available.  In 2003 we undertook
the design of a new prime-focus camera and wide-field
corrector in  order to provide WIRO with a modern, survey
imaging capability. The new  four-element corrector was
designed to produce sub-arcsecond images over the  40\arcmin\
field that the latest 4096 $\times$ 4096 CCDs would cover.
UV-grade  fused silica (Corning 7980 HPFS) was chosen for
the corrector material in  order to allow its use from the
atmospheric cutoff at $\sim$ 310 nm to the  long-wavelength
limit for silicon detectors, $\sim$ 1000 nm. The corrector 
design was  developed by Charles Harmer (NOAO) with the
assistance of one of us (HAK) using  the ZEMAX optical design
program. The RMS spot radius at wavelengths of 0.37,  0.55,
and 1.0 $\mu$m was minimized over fields of 0, 7, 14, 17,
and 20\arcmin\ along the X-Y diagonal. A back focal distance of 
$>$70 mm from the rear element was  maintained in order to
allow mechanical clearance for a 12 mm thick filter  wheel,
the mechanical focusing stage, and for the Dewar window to
safely clear the detector.

\begin{figure}
\centering
\includegraphics[width=\textwidth]{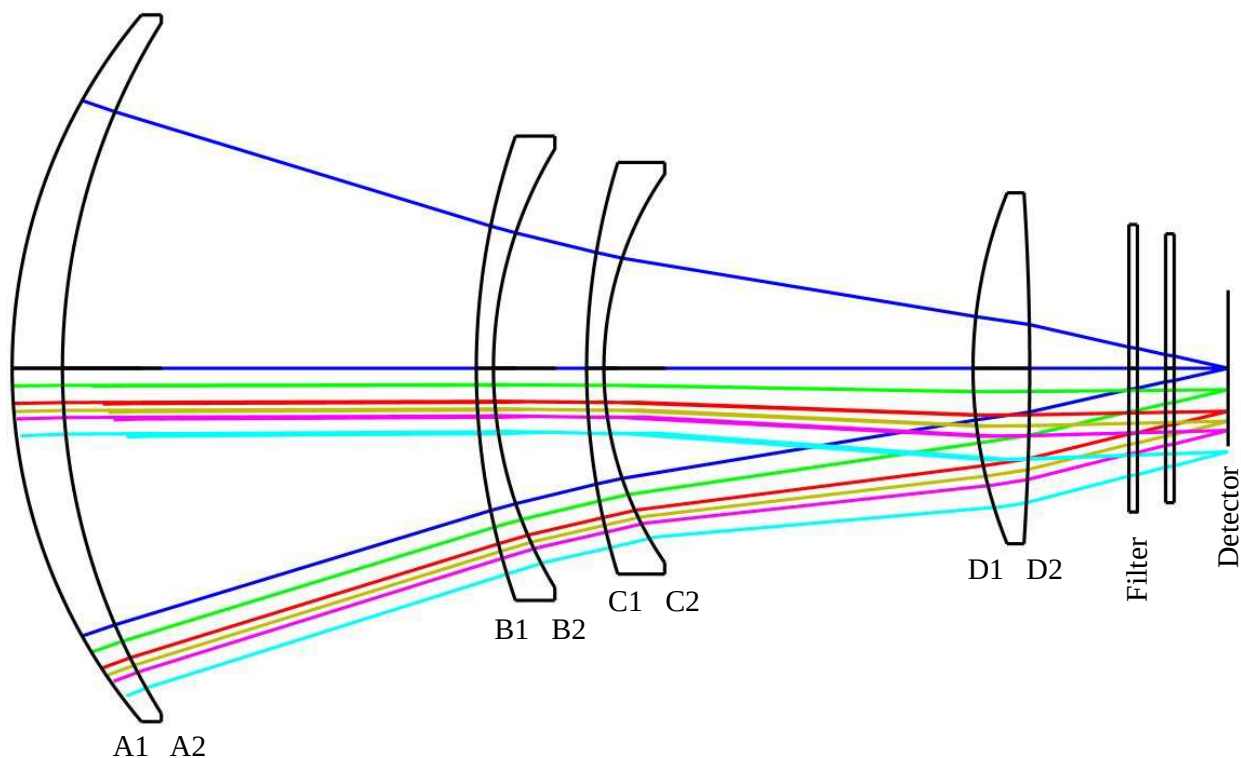}
\caption{Schematic design of the four-element prime-focus corrector
constructed for use on the WIRO telescope.  The diameter of the
largest element is 250 mm.  Labels 
correspond to the rows of Table~\ref{PFCelements.tab}, which
give the prescription for each surface.  
\label{pfc}}
\end{figure}

Figure~\ref{pfc} shows a schematic view of the corrector
with each of its eight spherical surfaces (A1, A2, B1, ...,
etc.) labeled.  Table~\ref{PFCelements.tab}
lists the prescription of each element. Each
lens surface (column 1)
entails a clear diameter (column 2), a thickness
and separation at center 
(column 3), and a radius of curvature (column 4). 
Tolerances are 0.1\% on each radius of curvature, 0.1 mm in
thickness, and $<$20 $\mu$m of thickness difference between
the two surfaces (wedge).  The glass mass of all four elements
is 3.0 kg.  Thermal analysis between 20\degr C and -20\degr C
indicated that aluminum was a suitable material for the lens
mounts and the entire optomechanical structure. 

\begin{deluxetable}{crrrr}
\centering
\tablewidth{0pc}
\tablecaption{Optical Prescription for the four-element WIRO 
Prime-Focus Corrector
\label{PFCelements.tab}}
\tablehead{
\colhead{Surface} & 
\colhead{Diameter } & 
\colhead{Thickness } &
\colhead{R$_{curve}$} \\
\colhead{} &
\colhead{(mm)} &
\colhead{(mm)} &
\colhead{(mm)}}
\tablecolumns{4}
\startdata  
A1 &  249.6 &  17.92 &  193.67 &  \\
A2 &  243.8 & 145.20 &  230.86 &  \\
B1 &  164.4 &   6.00 &  250.40 &  \\
B2 &  155.2 &  33.71 &  149.53 &  \\
C1 &  145.4 &   6.00 &  245.90 &  \\
C2 &  137.4 & 130.10 &  120.65 &  \\
D1 &  124.0 &  20.00 &  165.67 &  \\
D2 &  122.0 & \nodata& -949.12 &  \\
\enddata
\end{deluxetable}

\begin{figure}
\centering
\includegraphics[width=\textwidth]{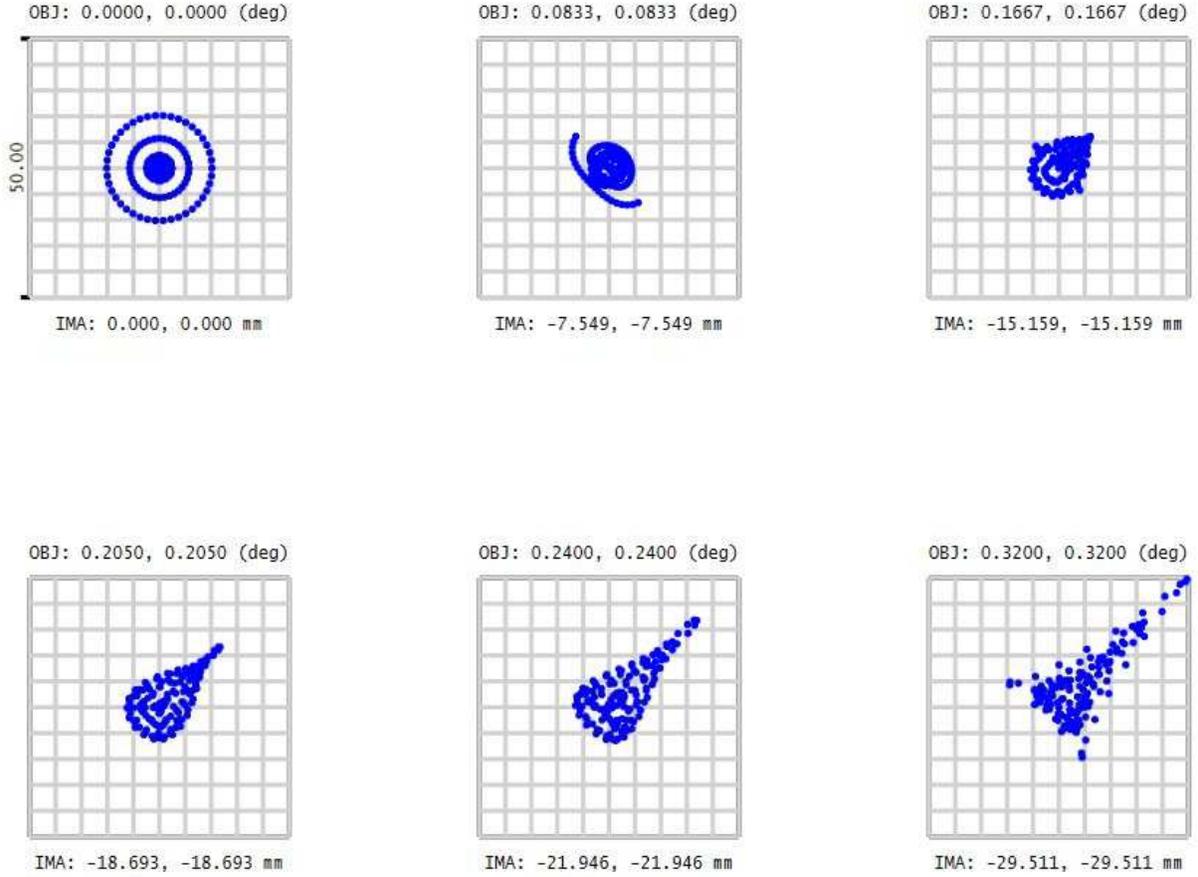}
\caption{Spot diagram illustrating imaging performance of the
WIRO prime-focus corrector as a function of field angle at 0.37 $\mu$m wavelength.  
Each box is 50 $\mu$m square, corresponding to approximately 2\farcs4 on the sky.  
Field angles of 0\arcmin, 5\arcmin, 10\arcmin,
12\arcmin, 14\arcmin, and 19\arcmin\ along the X-Y diagonal
are shown.  \label{pfcspot}}
\end{figure}

Figure~\ref{pfcspot} shows a spot diagram  illustrating the
designed optical performance of the corrector at 0.37 $\mu$m
wavelength at field angles of 0\arcmin, 5\arcmin,
10\arcmin, 12\arcmin, 14\arcmin, and 19\arcmin\ along the
X-Y diagonal.  The box is 50 $\mu$m square, corresponding
to  approximately 2\farcs4 on the sky, or about 4 pixels for
a 15 $\mu$m pixel size.  The  resulting spot sizes are
significantly smaller than 1\arcsec\ in all but the
largest field angles. The predicted 120 $\mu$m focus shift
between  wavelengths of 0.37 $\mu$m and 1 $\mu$m has proven
undetectable during  routine operation. The distortion over
the flat focal plane is less than 2\%.  The individual  lens
elements were manufactured by Harold Johnson Optical 
Laboratories, Inc.\ and coated with a single layer MgF$_2$
anti-reflection  coating optimized for 550 nm. 

The mechanical design of the corrector was subcontracted to
the Pilot Group  of Monrovia, CA.  Their design consists of
individual lens cells assembled and  pinned into a stack to
complete the corrector assembly. The design includes a 
system of thin, internal rings for stray light baffling.
The completed stack  of lens cells squeeze thin rubber
gaskets between each cell to seal the entire  assembly so
that it can be filled with dry N$_2$ in order to minimize
internal  condensation.  The
individual mechanical components were  manufactured by
Wyoming's Division of Research Support and flat black 
anodized in order to minimize internal reflections. 

The assembly and alignment process was done by one of us
(MP) on a precision  turntable via laser illumination of
each lens within its cell. Design specifications called for 
$<$1\arcmin\ tilt on each element, centering of the elements
to within $<$0.1 mm, and relative spacing of the elements to
within 0.1 mm. The procedure for aligning the 
lenses within their cells to ensure the optical axes correspond 
with the mechanical axis of the assembly is as follows;
each lens cell was placed upon the turntable, mechanically 
centered via a dial indicator and clamped in place. The lens
was then placed  upon its reference surface within its
oversized mount and the reflected laser  beam examined with
a microscope as the turntable was rotated. The 
laser does not need to be aligned with the rotational axis during this process. If 
the optical axis of the lens is aligned with the mechanical axis 
its beam will be reflected from a point at constant radius, and 
hence constant slope, as the lens cell is rotated. Thus the reflected 
beam does not move as the assembly is rotated. However if the optical axis 
of the lens is displaced from the mechanical axis a rotation of 
the assembly results in the beam reflecting off of a surface 
of changing slope and the beam will wobble as the assembly 
is rotated. The lens is adjusted until the wobble is removed 
at which point it reflects from a constant slope and radius 
meaning that the optical axis of the lens is now coincident 
with the mechanical axis of the assembly.

The position of 
the lens element was adjusted using four nylon radial
setscrews until the beam  was stationary indicating that the
optical axis of the lens was aligned with  the rotational
(mechanical) axis of its lens mount. Space grade silicone 
RTV (Dow Corning 93-500) was mixed with hardener and placed
within a vacuum  chamber in order to remove air bubbles. The
RTV was then drawn with a  large-gauge hypodermic needle and
extruded into the gap between the lens and  its  cell. When
the RTV had set, the adjustment screws were withdrawn and the
holes  filled with RTV. The lens and its mount were then set
aside and the process  repeated for the next lens and mount.
The corrector stack was assembled  using alignment pins and
screws to hold the individual lens cells and baffles  in
place while compressing a gasket between each pair of cells.
The completed  corrector assembly was then pumped free of
air and filled with dry N$_2$. Afterward, the ports were
filled with RTV. 

The corrector was then shipped to WIRO for incorporation
within the prime  focus camera assembly. The prime-focus
instrument package is  mounted on a large stage within a box
supported by  a four-vane spider of the prime-focus ring. 
The two axial stages  allow independent motion of the
corrector and the detector. The horizontal  instrument
stage allows the prime-focus assembly to be mechanically
centered  within the prime-focus ring of the 
telescope.\footnote{The corrector was originally installed and tested
with  the previous-generation prime-focus imager, known as
WIRO Prime, designed and built by MP. WIRO Prime was in
operation for approximately 10 years,  2004--2013.  WIRO
prime consisted of two stages mounted  on shafts using
linear bearings. Stepper motors turned precision screws to 
position the stages along WIRO's optical axis. A twelve slot
filter wheel holding two-inch filters 
and  the corrector were mounted to one of the
stages and an Astrophysical Research  Camera ND-5 Dewar
containing a 2048 $\times$ 2048 E2V CCD detector on the
other.  }   The corrector was aligned with the optical axis
of WIRO over a  period of several nights.  First, the
optical  axis of the telescope primary was identified by
imaging star fields without  the corrector in place. The
size of the strongly comatic star images were  measured and
the center of the pattern computed. The primary mirror was
then  aligned to bring its optical axis to the center of the
detector. The corrector installed at its
nominal axial position relative to the  detector. Images
taken inside and outside of focus were used to determine
the  position of best focus across the field and the
detector tilt computed. Shims  were used to ensure that the
detector mounted perpendicular to the optical  axis. 
Additional images were used to determine the optimal radial
and axial  location of the corrector, at which point the
corrector was judged to be  aligned. The
point-spread-function of the images across a 13\arcmin\
field of view were found to suffer from approximately 0.1\arcsec\
of astigmatism, but rotation of the corrector resulted
in no change of the position angle,  implying that the
astigmatism is present in WIRO's primary mirror. On the
best  nights at WIRO the resulting images have FWHM$\simeq$1.0\arcsec. 

\section{The WIRO Double-Prime Instrument Package}

In 2014 a new prime-focus instrument package  was
implemented at WIRO to take full advantage of the field
of view offered by the prime-focus corrector. A new focus
stage and filter wheel were constructed in the University of
Wyoming Division of Research Support shops.  A new 4k CCD camera was
purchased to enable wide-field imaging over the
$\sim$40\arcmin\ field provided by the corrector. Dubbed
``WIRO Double-Prime'' to signify the roughly 100\% increase
in field of view relative to the old prime-focus assembly
and 2k CCD camera, it was constructed during 2013--2014 and 
commissioned during the summer and fall of 2014.  

\subsection{Focus and Filter Wheel Mechanisms} 

\begin{figure}
\centering
\includegraphics[width=\textwidth]{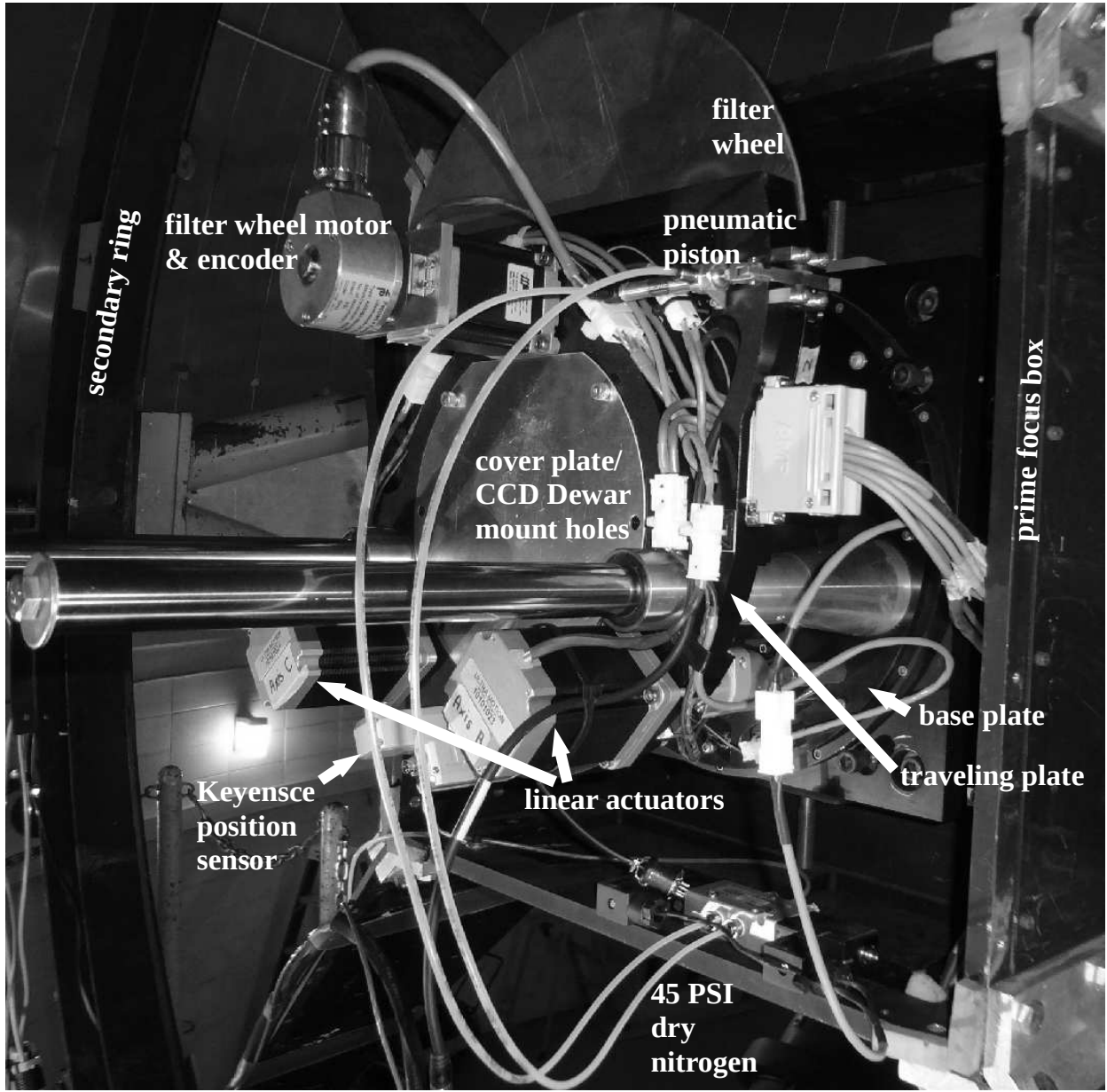}
\caption{Photograph of the WIRO Double Prime instrument package on the 
telescope, minus the CCD Dewar, which mounts at the location of the labeled 9-inch diameter 
cover plate.  
\label{primeunit}}
\end{figure}

Figure~\ref{primeunit} displays a labeled photograph of the WIRO
Double Prime prime-focus assembly. It is an upgraded version of
that used in WIRO Prime and it consists of a circular
aluminum base plate that mounts  to the telescope's
prime-focus interface box and a traveling plate that
supports the filter wheel, CCD Dewar, position sensor,  and
motors that drive the filter wheel and focus. The traveling
plate  translates along the optical axis via two parallel
stainless steel rails driven by two Ultramotion D1 linear
actuators  acting against a spring that supplies tension to
couple the traveling plate to the base plate. The actuators
are coupled in software to ensure simultaneous motion.   The
distance between the base plate and traveling plate is
encoded to within $\sim$1 $\mu$m rms by a Keyence GT2
Digital Contact Sensor.   The traveling plate supports a
five-position filter wheel that can accept filters up
to 3.25 inches square; smaller filters may be used with the
addition of custom insets, at the cost of  reduced field of
view.   A pneumatic piston driven by a 45 p.s.i.\ dry
nitrogen supply hose locks the wheel at one of the five
filter positions defined by the location of conical detents
machined into the periphery of the filter wheel.  This
passive mechanical torque obviates  the need for the Applied
Motion Products Inc.\ high-torque NEMA size 23 stepper motor
to supply holding torque on the filter wheel and ensures
precise filter positioning.  A Pepperl+Fuchs AHS58-H
absolute single-turn encoder with 16-bit resolution located
on the shaft of the filter wheel motor monitors filter wheel
rotation to 0.33\arcmin. A Dewar containing the CCD
mounts at the location of the cover plate pictured in 
Figure~\ref{primeunit}.  The Uniblitz\footnote{Uniblitz is now part of Vincent Associates.}
model CS90HS1T0 iris type shutter, 
driven by a Uniblitz model VCM-D1 driver, resides
just behind the cover plate.  
A Galil DMC-4040 motion controller drives the
focus and filter wheel motors.  Motion commands are
implemented within a custom LabVIEW application with a
graphical user interface running on a linux operating system. 
Real-time focus and filter position information are
displayed to the user and recorded for inclusion in the
FITS image headers \citep{Wells81, Pence10}.    

\subsection{CCD Camera} 

The WIRO Double-Prime imaging device is a 4096$\times$4096
thinned back-illuminated CCD fabricated at the University of Arizona
Imaging Technology Laboratory.  Enhanced UV coatings
optimize  the quantum efficiency for the 3600 -- 4500 \AA\
portion of the spectrum.  Actual peak quantum efficiency
exceeds 85\% at 3900 \AA\ and remains above 75\% through
7000 \AA. The CCD was mounted in a custom Dewar fabricated
by Astronomical Research Cameras, Inc., which also supplied
the readout electronics (ARC Generation III controller).  
The CCD window is fabricated from UV-grade fused silica to
preserve high blue throughput.

\subsection{Detector Characteristics}

\begin{figure}
\centering
\includegraphics[width=\textwidth]{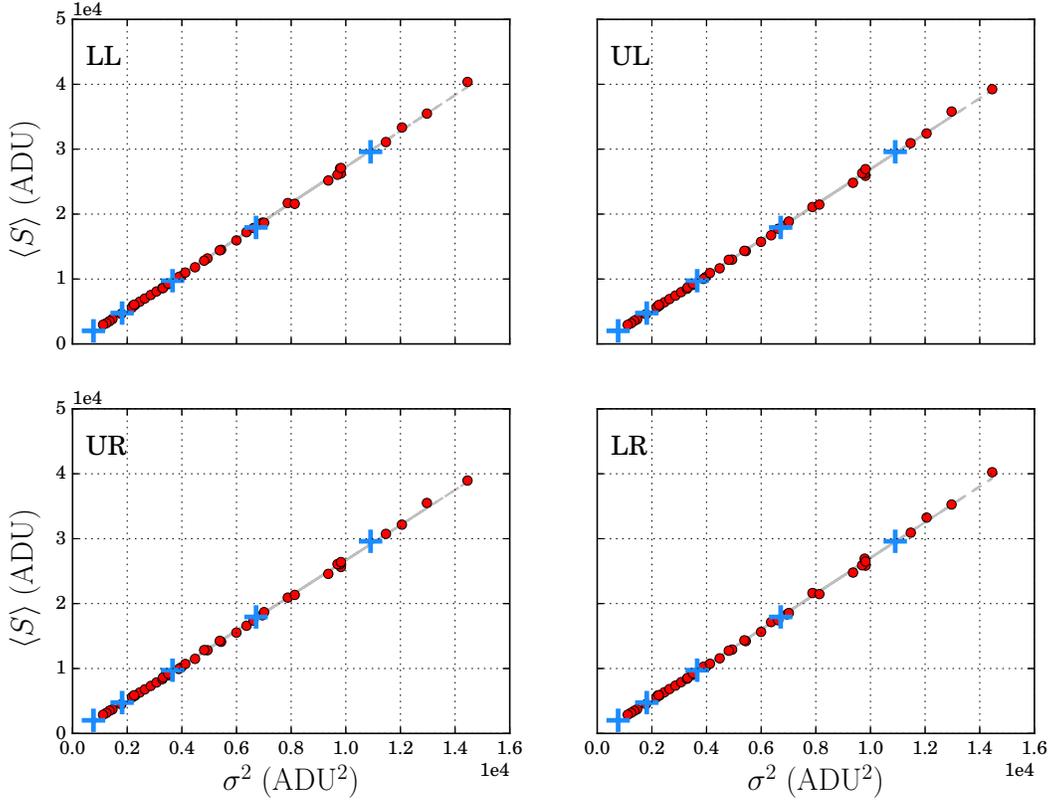}
\caption{Signal-versus-variance $(\langle S \rangle - \sigma^2)$ plots for each of the
four quadrants of the Double-Prime array. Clockwise from the top left, panels
correspond to the lower-left (LL$\equiv$north-east), upper-left (UL$\equiv$north-west), 
lower-right (LR$\equiv$south-east) and 
upper-right (UR$\equiv$south-west) quadrants respectively. The signal was measured in clean regions
of each quadrant over pairs of three-second dome-flats at increasing lamp intensity
levels. The variance was measured in the difference images between pairs of 
dome flats in order to remove pixel-to-pixel variation in the flat field.
The gray lines are the best fits to the red filed circles, which are the data taken from 
the set of three-second dome flats. The slope of the line gives the
gain of the relevant CCD (see Table~\ref{gainread.tab}). The blue crosses show the signal and
variance measured from a set of ten-second dome flats. The resulting gain measurments are entirely
consistent with the results found from the three-second flats. 
\label{gain}}
\end{figure}

The detector gain was determined for the default 
four-amplifier readout mode from the slopes of the
signal-versus-variance relations shown in Figure~\ref{gain}.
Both quantities  were measured from pairs of three-second
dome flats   with consecutive pairs taken at  increasing
dome lamp intensity levels. The results are shown by the
red filled circles in Figure~\ref{gain} and the best linear 
fits to these measurements are shown by the gray lines.
To monitor the  stability of the
light source, pairs of exposures were  interleaved by a
single two-second exposure at a constant  intensity level. 

The plotted quantity, $\langle S \rangle$, is the average 
signal in ADU measured over each image pair in clean regions
of  each quadrant of the detector array. The variance,
$\sigma^2$,  was measured over identical pixel regions in
the difference image between each image pair.\footnote{Note
that the plotted quantity $\sigma^2$ is  half the variance
measured in a given difference image. The  extra factor 0.5
accounts for the fact that the variance is  doubled when
subtracting two dome flat pairs.} This process reduces the
effect of pixel-to-pixel sensitivity  variations in the
flat  field which would otherwise produce scatter  in the
pixel levels and would lead to a biased gain  measurement.
The procedure does not account for noise introduced  during
bias subtraction of the flat field frames, however for  the
low level Poisson distributed counts in bias frames these
noise terms are negligible. Table~\ref{gainread.tab} lists
the  gain measured in each quadrant from the best linear fit
to the  signal-to-variance relation over all 41 dome flat
pairs. The random measurement errors are calculated from the scatter in the
relation after  iteratively re-fitting with a single data
point omitted at  each step.  Measured gains are very near 2.6
electrons ADU$^{-1}$ in all quadrants.

The use of three-second flats in the gain measurements was chosen
to ensure dark current remained negligible throughout. To check that these
short exposures do not suffer from shutter timing effects (see below), which
could otherwise bias our gain measurements, the procedure was repeated with
a set of ten-second dome flats. The resulting measurements are overlaid in
Figure~\ref{gain} as blue crosses and are found to be entirely consistent with
the results returned from the three-second flats.
 
Using the measured gain, $G$, the readnoise in electrons 
was calculated from the standard deviation in the 
difference images of bias frame pairs, $\sigma_{B1-B2}$, 
as $\sigma_{RN}=G\frac{\sigma_{B1-B2}}{\sqrt{2}}$. The 
readnoise calculations were performed over identical 
pixel areas as defined for the gain measurements, and 
the quantities reported in Table~\ref{gainread.tab} are 
the median readnoise measurements over 342 unique bias 
frame pairs for each quadrant. 

\begin{deluxetable}{ccc}
\centering
\tablewidth{0pc}
\tablecaption{WIRO Double Prime Gain and Readnoise
\label{gainread.tab}}
\tablehead{
\colhead{Detector} &
\colhead{Gain} &
\colhead{Readnoise} \\
\colhead{ } &
\colhead{ $(\mathrm{e^-/ADU})$ } &
\colhead{ $(\mathrm{ADU})$ }}
\tablecolumns{3}
\startdata  
LL (NE) & $2.677 \pm 0.003$ & $5.41 \pm 0.16$\\
UL (NW) & $2.639 \pm 0.003$ & $5.17 \pm 0.21$\\
UR (SW) & $2.619 \pm 0.003$ & $5.20 \pm 0.21$\\
LR (SE) & $2.664 \pm 0.004$ & $5.19 \pm 0.21$\\
\enddata
\end{deluxetable}

\begin{figure}
\centering
\includegraphics[width=\textwidth]{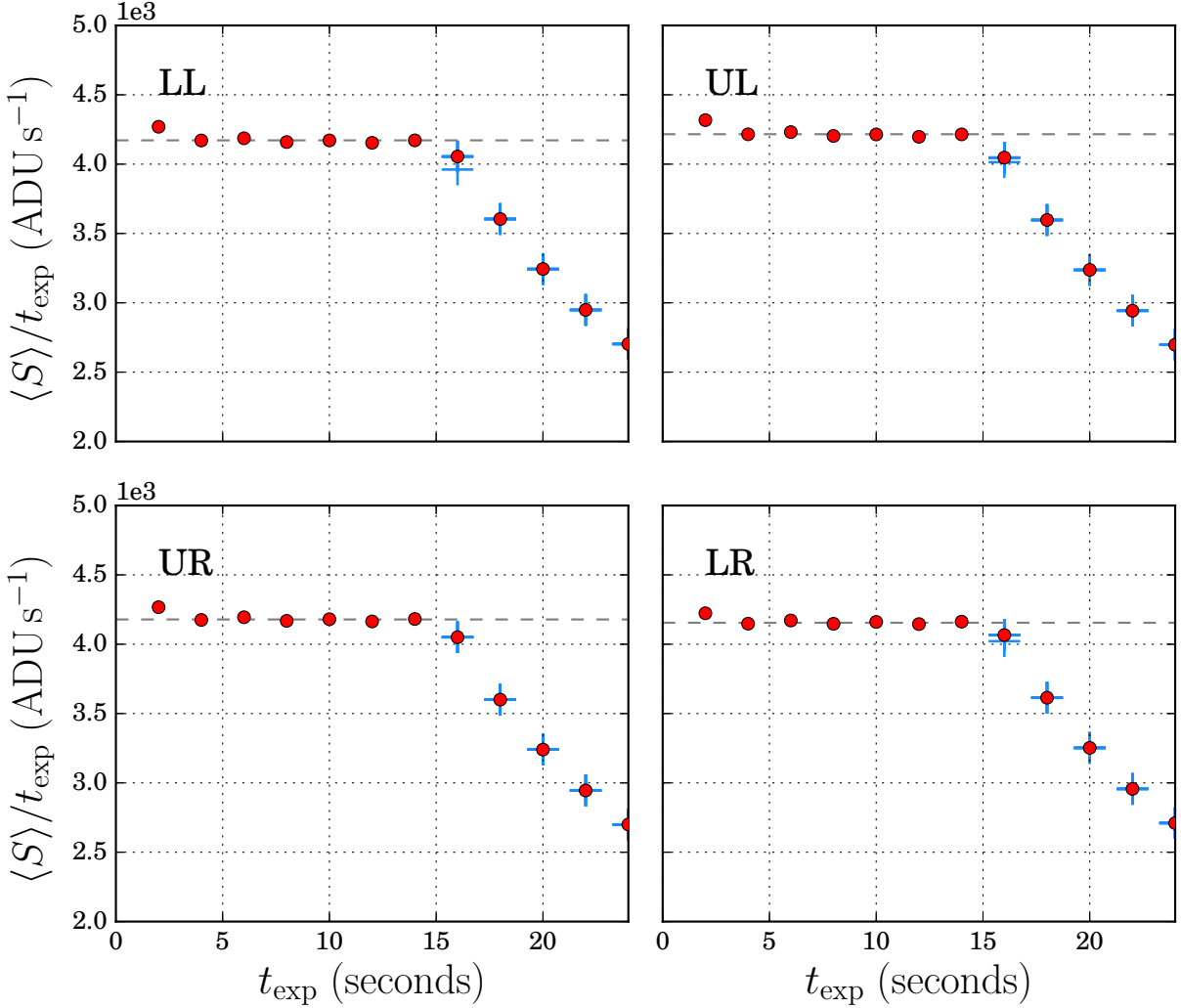}
\caption{Linearity curves for each of the four Double-Prime array quadrants. Clockwise 
from the top left, panels correspond to the lower-left (LL$\equiv$north-east), 
upper-left (UL$\equiv$north-west), 
lower-right (LR$\equiv$south-east) and upper-right (UR$\equiv$south-west) quadrants respectively. Signal was measured 
in a series of dome flats at constant intensity with exposure times increasing in 
two-second intervals until saturation. Filled circles give the average count rate in units of 
$\mathrm{ADU\, s}^{-1}$. Five crosses at each interval beyond saturation give the average count rate in small 
regions sampling the corners and center of each CCD. At each interval crosses are 
largely indistinguishable from one another on the plot scale. Each detector saturates at
$\sim 60000\ \mathrm{ADU}$ and there is no measurable difference in the saturation 
level between CCDs or within any single CCD within counting errors. The dashed
gray line shows a linear fit with forced zero slope to data points between 
$t_\mathrm{exp} = (4 - 14)\,\mathrm{s}$. The deviation to the fit at $t_\mathrm{exp} = 2\,\mathrm{s}$
is likely attributable to shutter timing effects.
\label{linearity}}
\end{figure}

The linearity of all four quadrants has been characterized 
via a sequence of dome flats at constant light level exposed
for durations  increasing by two seconds until  saturation.
Each integration was followed by an   exposure of two
seconds in order to monitor the stability  of the light
source.  Figure~\ref{linearity} illustrates the 
linearity curves plotted as the average count rate versus time 
for each CCD quadrant. Saturation occurs
broadly at $\sim \mathrm{60,000\, ADU}$ across all detectors
with no measurable difference in the  saturation level in
each quadrant within counting errors. The  blue crosses
correspond to the count rate measured in five  smaller
regions sampling the corners and center of each  quadrant.
The five data points at each interval are largely indistinguishable
on the plot scale and there is no indication of a measurable variation 
in the saturation level across each detector within  counting 
errors. The dashed gray line is a linear fit with forced zero slope
to the data between $t_\mathrm{exp} = (4 - 14)\,\mathrm{s}$. There is
an obvious deviation from the fit at $t_\mathrm{exp} = 2\,\mathrm{s}$.
Although we do not rule out the possibility of non-linearity 
at low signal levels the 
more likely origin of this effect is the finite shutter open/close timing, 
which will alter both the true exposure time and the illumination
pattern across the array when the duration of the exposure is comparable 
to the shutter timing.

\begin{figure}
\centering
\includegraphics[width=\textwidth]{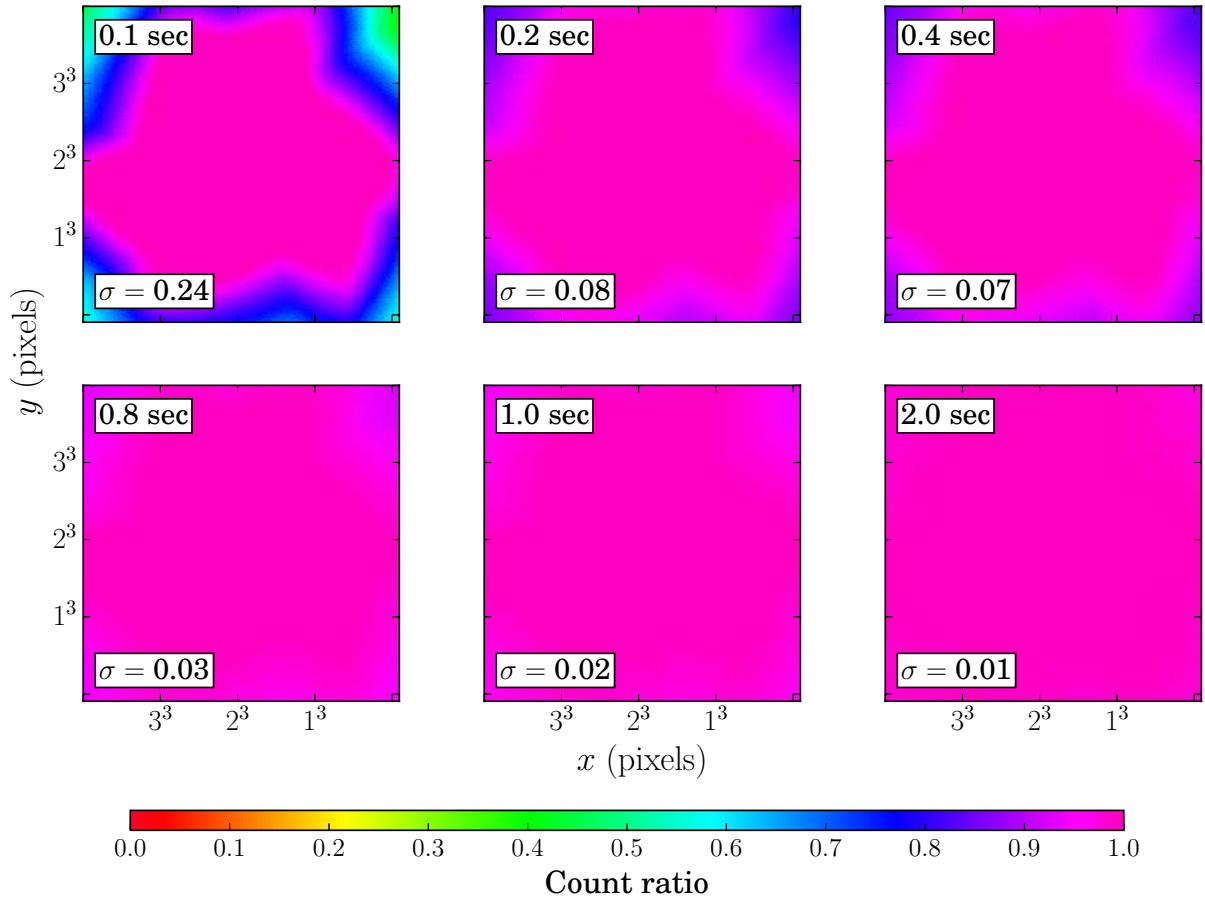}
\caption{Shutter timing effects shown as the ratio of a series of dome flats 
to a reference eight-second dome flat at a constant light intensity level. Plot legends 
give exposure times in the top left corners and standard deviations in the bottom left
corners for the relevant flat ratio in the series. Non-uniform illumination of the array is clear during
the shortest exposure times but undetectable in the two-second flat.
\label{shutter}}
\end{figure}

The effects of shutter timing on illumination have been investigated using a 
series of dome flats taken at constant light levels for exposure 
times varying between 0.1 and four seconds. Figure~\ref{shutter} 
shows the ratio of a sample of these dome flats to an 8 second
dome flat after normalizing by exposure time. The legends in each
panel give the exposure time of the comparison image in the top
left and the standard deviation of the flat ratio in the bottom left. 
Illumination clearly varies across the detector arrays 
at short exposure times, with the edges of the fields most affected. 
As the exposure time increases the effect becomes less 
pronounced until the illumination across a two-second flat-field 
becomes indistinguishable from that across an eight-second 
flat-field. Due to the effects identified in
Figures~\ref{linearity} and~\ref{shutter}, exposure times of 
less than four seconds are not recommended.

\subsection{Image Quality}\label{sebsec:imgquality}

\begin{figure}
\centering
\includegraphics[width=\textwidth]{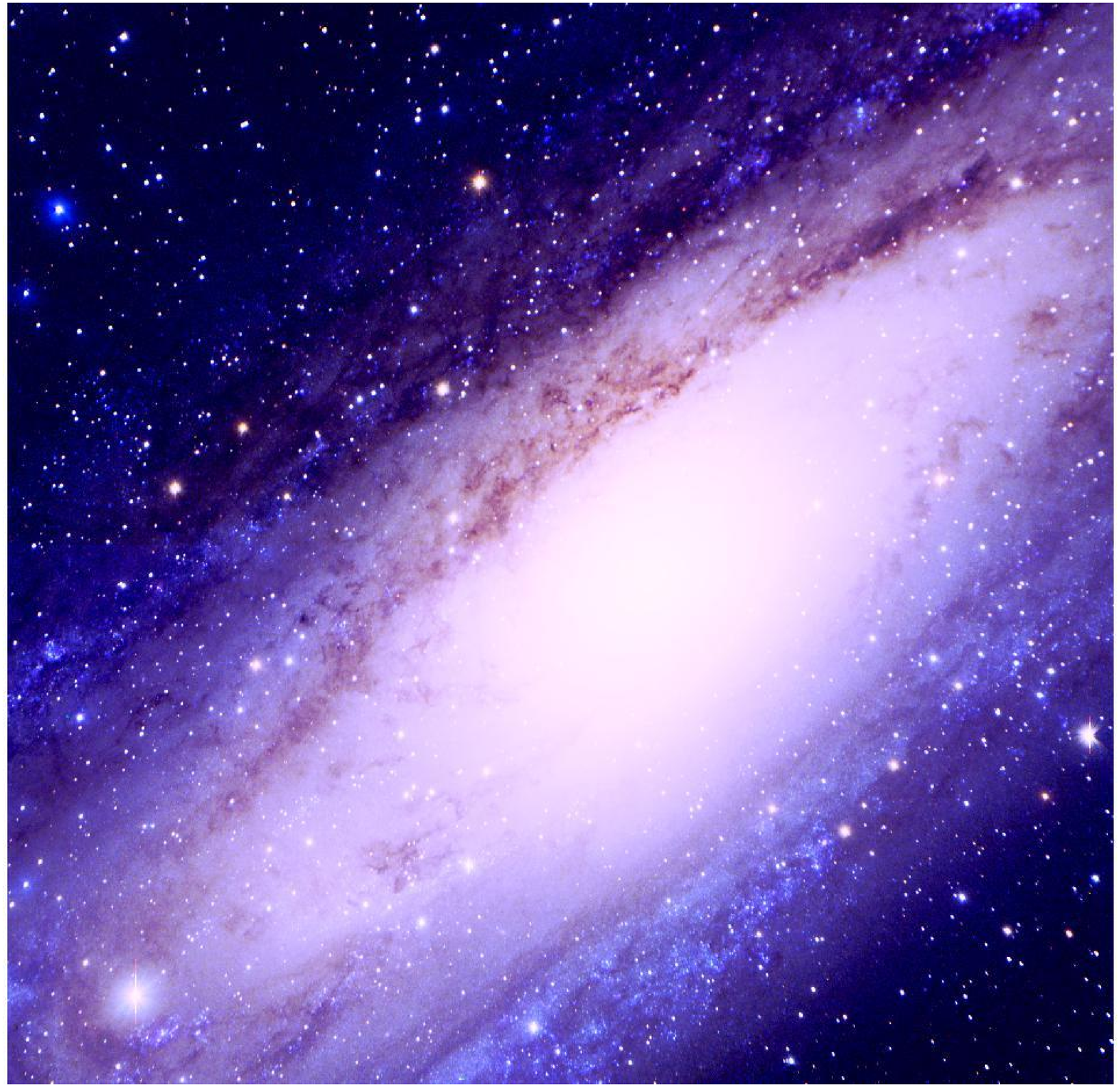}
\caption{A $u^{\prime}/g^{\prime}/r^{\prime}$ image of M31, illustrating the 
nearly 40\arcmin\ field of view. 
\label{m31}}
\end{figure}

WIRO Double-Prime has a field of view of approximately 
39\arcmin\ at a nominal on-axis scale of 0.58\arcsec\
pix$^{-1}$.  For comparison, the full moon subtends an
average angle of  31\arcmin\ on the sky. Figure~\ref{m31}
shows a false-color  image of M31, produced from a
combination of three 60 s exposures in the
$u^{\prime}$/$g^{\prime}$/$r^{\prime}$ filters, represented
as blue/green/red.\footnote{Some residual off-axis
aberrations can be  seen at large  field angles as a  result
of non-optimal corrector alignment that is still being
diagnosed as of this writing.}    The instrument suffers
from a small amount of pincushion  distortion, which is
inherent in the design of the  four-element prime-focus
corrector. The nominal plate scale  at the center of the
field is 0\farcs581 pix$^{-1}$ but  drops to near 0\farcs566
pix$^{-1}$  near the corners of  the usable field
(27\arcmin\ field angles).  A geometrical transformation 
(such as the {\sl IRAF} {\tt geotran} task, or similar), 
determined by mapping a bright star across a regular grid 
pattern, is used to correct the images to a uniform plate 
scale of 0\farcs58 pix$^{-1}$ prior to co-adding dithered 
images or performing other high-level processing steps.

\subsection{Filter Set, Sensitivity \& Color terms}

\begin{figure}
\centering
\includegraphics[width=\textwidth]{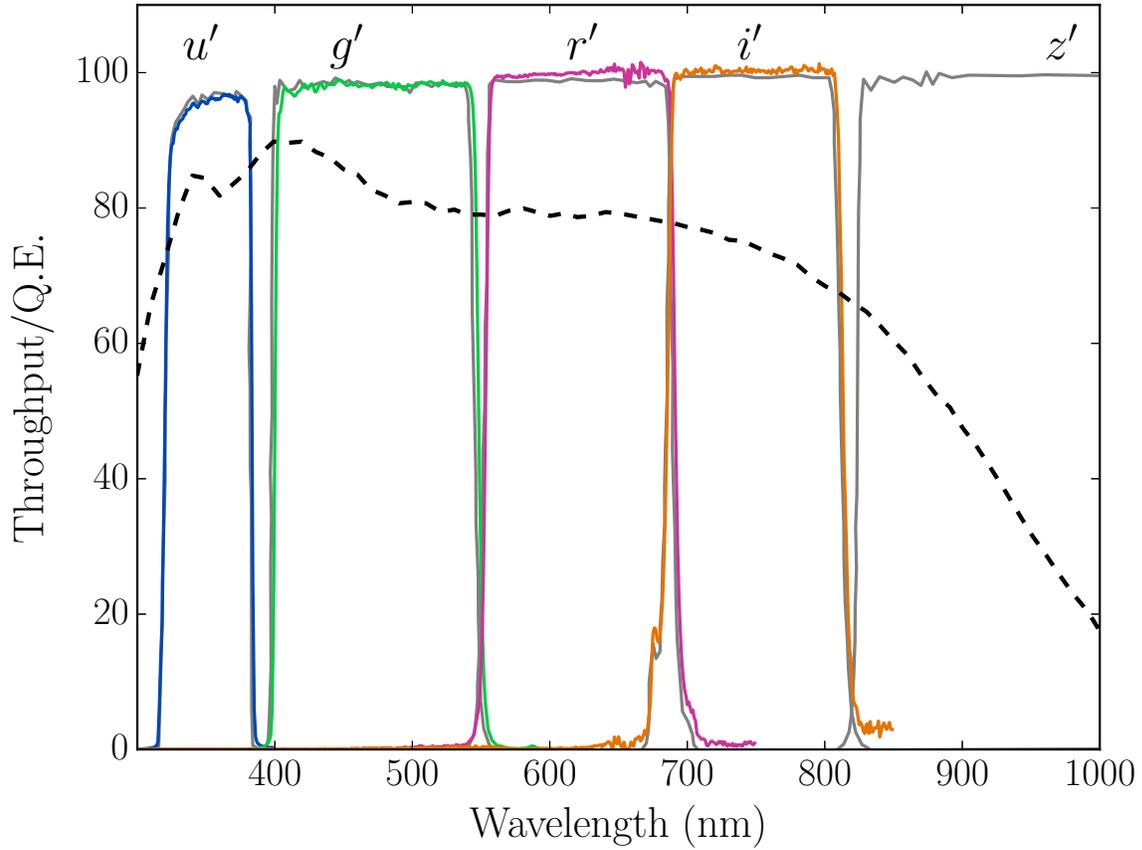}
\caption{The measured throughput of the Astrodon $u^{\prime}, 
g^{\prime}, r^{\prime}$, and $i^{\prime}$ filters (colored curves).
The $z^{\prime}$ band filter curve has not been accurately 
measured by the authors at the time of writing.
Gray curves show the manufacturer archetype filter curves, 
which differ from the
true filter curves by no more than a few per cent. 
The black curve shows the average quantum 
efficiency of the Double-Prime 4096$^{2}$ array. 
\label{filters}}
\end{figure}

Under normal operation the filter wheel houses five
Astrodon  manufactured SDSS-like filters \citep{Doi10}
denoted  $u^{\prime}, g^{\prime}, r^{\prime}, i^{\prime},
z^{\prime}$.  Figure~\ref{filters} plots the measured filter
curves for $u^{\prime}, g^{\prime}, r^{\prime}$ and
$i^{\prime}$.\footnote{The $z^{\prime}$  band filter curve
has not been accurately measured by the authors at the time of writing.} 
The manufacturer's archetype filter curves  are plotted in
gray and differ from the true curves by no  more than a few
percent. Also plotted in Figure~\ref{filters}  is the
average detector quantum efficiency curve  (provided
courtesy of R. Leach).

Ideally one would like to characterize the sensitivity of
the  telescope and optical system by tying Double-Prime
imaging data  to the ``natural magnitude system'' defined by
wavelength-dependent  quantities such as the photon weighted
central wavelengths of each  filter, the quantum efficiency
of the CCD array, the mirror  reflectively and the
characteristic atmospheric absorption at  Mt.\ Jelm. Since
the total system throughput is not well known we  instead
characterize the sensitivity of the telescope with respect 
to the SDSS magnitude system.

SDSS photometry is intended to be on the AB system
\citep{OkeGunn83},  however small offsets from the AB
standard are known to exist \citep{Smith02}. 
Nevertheless the wide use of the SDSS system throughout
astronomy and  the close relationship between the SDSS and
Double Prime filter sets (see below) make it a good choice
on which to base the following analysis.

The system sensitivity has been characterized from a single
night's observations of a field at \ra=2 hrs, \dec=+1\arcdeg 
(within the SDSS Stripe 82 region \citep{Annis14}) at airmasses
ranging from between  1.27 to 2.5. The data were taken in
photometric conditions during which  the average
measured extinction was $\kappa(u^{\prime}, g^{\prime}, 
r^{\prime}, i^{\prime}, z^{\prime}) = (0.47, 0.17, 0.09, 0.07, 0.06)
$ mag airmass$^{-1}$. The Double-Prime zero points for the night are given in 
Table~\ref{zeropoints.tab} as the average difference between
the measured stellar instrumental magnitudes and the SDSS
magnitudes  over the target field linearly extrapolated to
zero airmass. It  should be stressed that even in
photometric conditions zero points  will vary on a nightly
basis due to changing atmospheric conditions  and over
longer time scales due to wide ranging factors such as 
degradation of the mirror coating or the anti-reflection
coatings on  optical elements. Therefore these results
should serve as a guide only,  useful for comparisons with
other telescopes or for approximate ``on the mountain''
photometry.

\begin{deluxetable}{cr}
\centering
\tablewidth{0pc}
\tablecaption{WIRO Double Prime Zero Points
\label{zeropoints.tab}}
\tablehead{
\colhead{Band} & 
\colhead{Z.P.}  \\
\colhead{ } &
\colhead{(mag.)}}
\tablecolumns{2}
\startdata  
$u^{\prime}$ (Astrodon) &  $24.52 \pm 0.03$ \\
$g^{\prime}$ (Astrodon) &  $25.31 \pm 0.02$ \\
$r^{\prime}$ (Astrodon) &  $24.90 \pm 0.02$ \\
$i^{\prime}$ (Astrodon) &  $24.37 \pm 0.02$ \\
$z^{\prime}$ (Astrodon) &  $23.74 \pm 0.03$ \\
\enddata
\end{deluxetable}

\begin{figure}
\centering
\includegraphics[width=\textwidth]{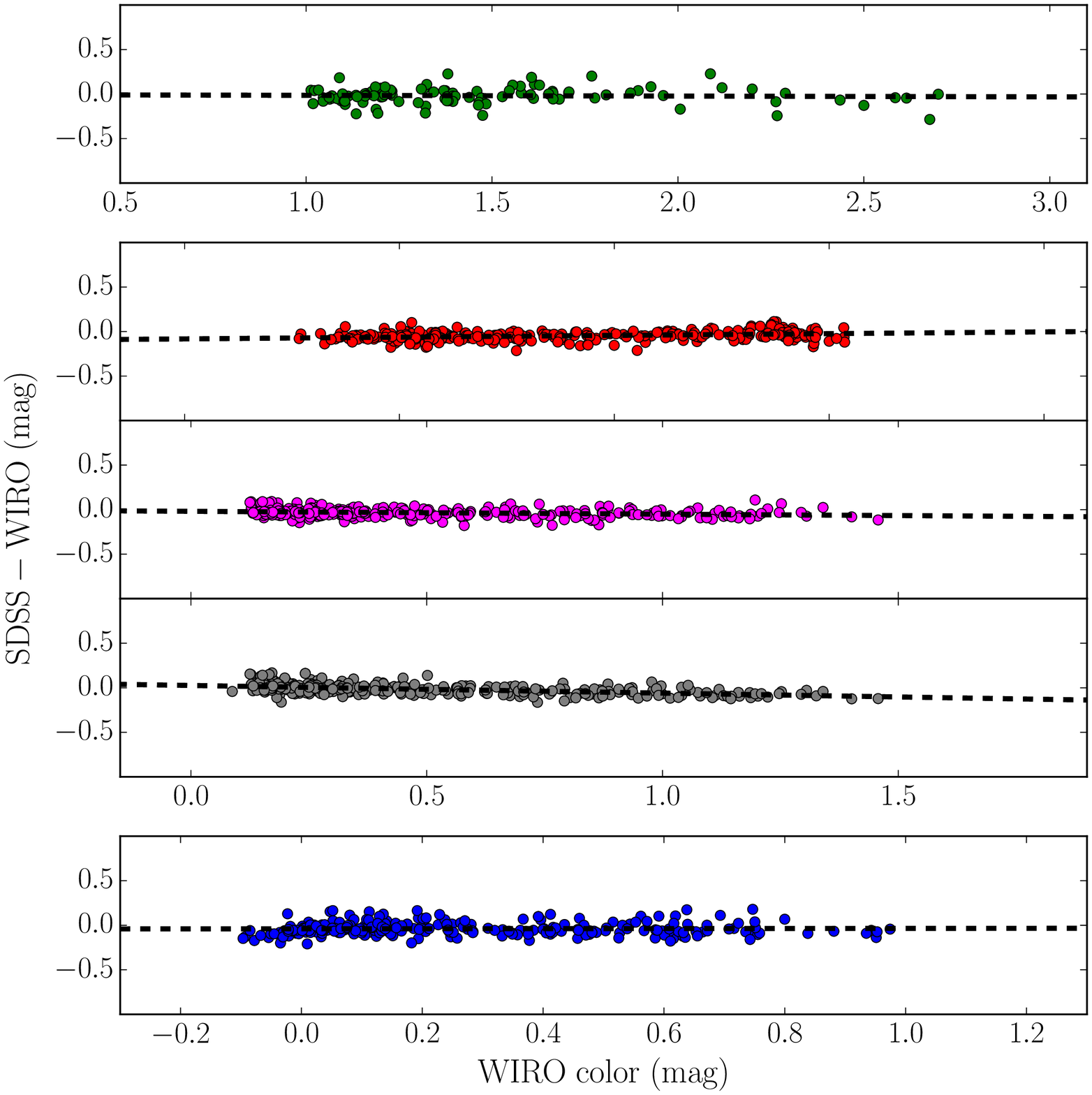}
\caption{Color relationships between WIRO Double Prime and SDSS imaging in the Astrodon
$(u^{\prime},\ g^{\prime},\ r^{\prime},\ i^{\prime},\ z^{\prime})$ and standard SDSS $(u,\ g,\ r,\ i,\ z)$
filter sets. Top to bottom shows $u-u^{\prime}:u^{\prime}-g^{\prime}$, $g-g^{\prime}:g^{\prime}-r^{\prime}$, $r-r^{\prime}:r^{\prime}-i^{\prime}$, $i-i^{\prime}:r^{\prime}-i^{\prime}$,
$z-z^{\prime}:i^{\prime}-z^{\prime}$. The dashed lines show the best fit to the main distribution of stellar colors
plotted as filled circles. The best fit color relations are
given by equation set~\ref{eqn:colors}. \label{colors}}
\end{figure}

In Figure~\ref{colors} we compare the measured Double Prime
and SDSS  archive colors of stars in the target field to
define color transformations between the two systems. We plot
SDSS PSF magnitudes versus Double Prime aperture magnitudes measured in
a wide 16-pixel radius aperture to reduce the need for  significant
aperture corrections, which can be difficult to determine due to
image deformation at the field edges (see
section~\ref{sebsec:imgquality}).  From top to bottom
Figure~\ref{colors} plots $u-u^{\prime}:u^{\prime}-g^{\prime}$, 
$g-g^{\prime}:g^{\prime}-r^{\prime}$, $r-r^{\prime}:r^{\prime}-i^{\prime}$, $i-i^{\prime}:r^{\prime}-i^{\prime}$
and $z-z^{\prime}:i^{\prime}-z^{\prime}$.  In each panel the best fitting
linear color relation is shown as a  dashed line. The fit
was made to all objects in the field with magnitude errors $\delta(\mathrm{m}) < 0.05$ 
after iteratively clipping $3\sigma$  outliers.
It is clear from the plot that the color terms in each case are small. 
In comparison
to the SDSS filters, the Astrodon filters have broadly similar central wavelengths but their
overall shapes are more top-hat like. The Astrodon $r^{\prime}$ band is slightly broader than the 
corresponding SDSS band while the SDSS $g$ and $i$ bands are shifted slightly to the blue and 
red respectively compared to $g^{\prime}$ and $i^{\prime}$. These differences likely account for the 
the comparatively larger color terms in these bands. The best fit color relations are defined as follows;

\begin{align}\label{eqn:colors}
u&=u^{\prime}-0.008(u^{\prime}-g^{\prime})\nonumber\\
g&=g^{\prime}+0.039(g^{\prime}-r^{\prime})\nonumber\\
r&=r^{\prime}-0.031(r^{\prime}-i^{\prime})\nonumber\\
i&=i^{\prime}-0.086(r^{\prime}-i^{\prime})\nonumber\\
z&=z^{\prime}+0.004(i^{\prime}-z^{\prime})\nonumber\\
\end{align}

\section{Remote Operation at WIRO}
\subsection{Infrastructure for Fully Remote Operation}

Good astronomical observing sites are invariably remote and can present 
significant challenges to observatory operations and users alike. Since the arrival of
high-bandwidth low latency networking in the mid 1990s professional 
observatories have realized the increase in efficiency, cost effectiveness
and safety offered through remote operations. Observatories can now be
minimally staffed, users do not need to travel long distances, coordination 
of food and water supplies are less complex, operations can continue during 
periods of hazardous weather, etc. Consequently most professional observatories
now offer full telescope and instrument control from somewhere other than a 
mountain top. 

The 10\,m Keck telescopes have offered remote operations for many of 
their instruments from their headquarters in Waimea since 1996 and 
from the United States mainland since 1998 \citep{Wirth08}. 
Keck has shown that solutions can be both adaptable and scalable by
serving facilities with similar instrument control interface designs 
such as those at the UCO/Lick Observatory including the 1\,m Nickel 
\citep{Grigsby08}. Other professional class optical and near infrared
telescopes providing detailed literature on their remote 
capabilities include Subaru \citep{Kosugi04}, the Canada-France-Hawaii 
Telescope \citep[CFHT;][]{Vermeulen93}, the NASA Infrared Telescope 
Facility (IRTF) \citep{Bus02} and the Southern Astrophysical Research 
Telescope \citep[SOAR;][]{Cecil02}.  

As networking and remote control technology has progressed, the technical 
infrastructure costs have fallen, allowing University owned observatories 
to take on similar projects with small telescopes. As well as enjoying the
benefits of increased operational efficiency these smaller observatories are
often able to increase student collaboration and training. Recent examples
include the Antarctic Bright Star Survey Telescope \citep[BSST;][]{Zhang16},
the Lee Sang Gak Telescope \citep[LSGT;][]{Im15} and the Lowell Observatory
0.8\,m Telescope \citep{Bui10}. 

In realizing the fully remote operation of
WIRO the authors undertook a slightly different challenge; WIRO is neither 
a large national facility nor a small educational project. WIRO has power,
logistical and operational demands that are comparable to a small national 
facility but has just two dedicated full time staff.  
Since its construction WIRO has been operated in the
traditional manner, requiring at least two on-site observers as a
matter of safety in light of its remote location.  Students and
faculty serve as the instrument scientist, telescope
operator, and observer, given that there are no nighttime
staff.  In 2010 we committed to a series of infrastructure 
upgrades to enable single-user operation from the safety and
convenience of the UW campus. Fully remote operation 
required installation of reliable backup power, reliable
high-bandwidth communications, and a series of modifications
to observatory infrastructure to enable remote operation of 
electrical circuits and key facility mechanisms used during
nightly observing.   

A CumminsOnan Model 60DSFAD diesel autostart generator rated
to 55 kW  at 9000 feet elevation was installed in the fall
of 2013, replacing the obsolete 1970's-era 8 kW generator.  
Failure of line power triggers automatic activation of the
generator and  the Cummins model OTPCB  transfer switches that disconnect
line power and enables the generator feed.  A delay of two
minutes is imposed to prevent unnecessary generator wear
during momentary power events. The generator is housed in a
heated ($\gtrsim$3\degr\ C) machine shed attached to the
northeast exterior wall of the dome. The 140 gallon diesel
fuel tank is capable of  supplying power sufficient for all
observatory functions for up to seven days.  

Critical observatory functions are powered by a 10 kVA
uninterruptible power supply (UPS) with a runtime of 
$\sim$12 minutes at nominal nighttime observing loads. 
A network interface within the UPS is configured to notify staff
via email or text upon power outages or critical events. 
UPS-powered items include the telescope drive motors and
control system, dome shutters, mirror covers, dome video
cameras,  network infrastructure,  telescope control computer,
and data acquisition computers.  The
nominal run time at full load is sufficient to allow startup
of the diesel generator or, failing that, safe shutdown of
the observatory.  

Observatory control communications occur over
ethernet using a dedicated local area network (LAN) at WIRO
anchored by a 48-port network switch.   A pair of
Motorola  PTP800 transceivers and 2' diameter antennas
provides a 25-mile line-of-sight link between
WIRO and the UW campus network.  Radios at each end of the
link have their independent battery backup with a 12-hour
runtime. The licensed 11 GHz channel provides throughput of
300 megabits s$^{-1}$, sufficient for  routine operation that requires
remote displays (via Vncviewer remote desktop) of WIRO computer desktops,
streaming video from dome cameras, monitoring of critical
observatory functions, and periodic transfer of astronomical
images (32 MB per exposure).   

\begin{figure}
\centering
\includegraphics[width=\textwidth]{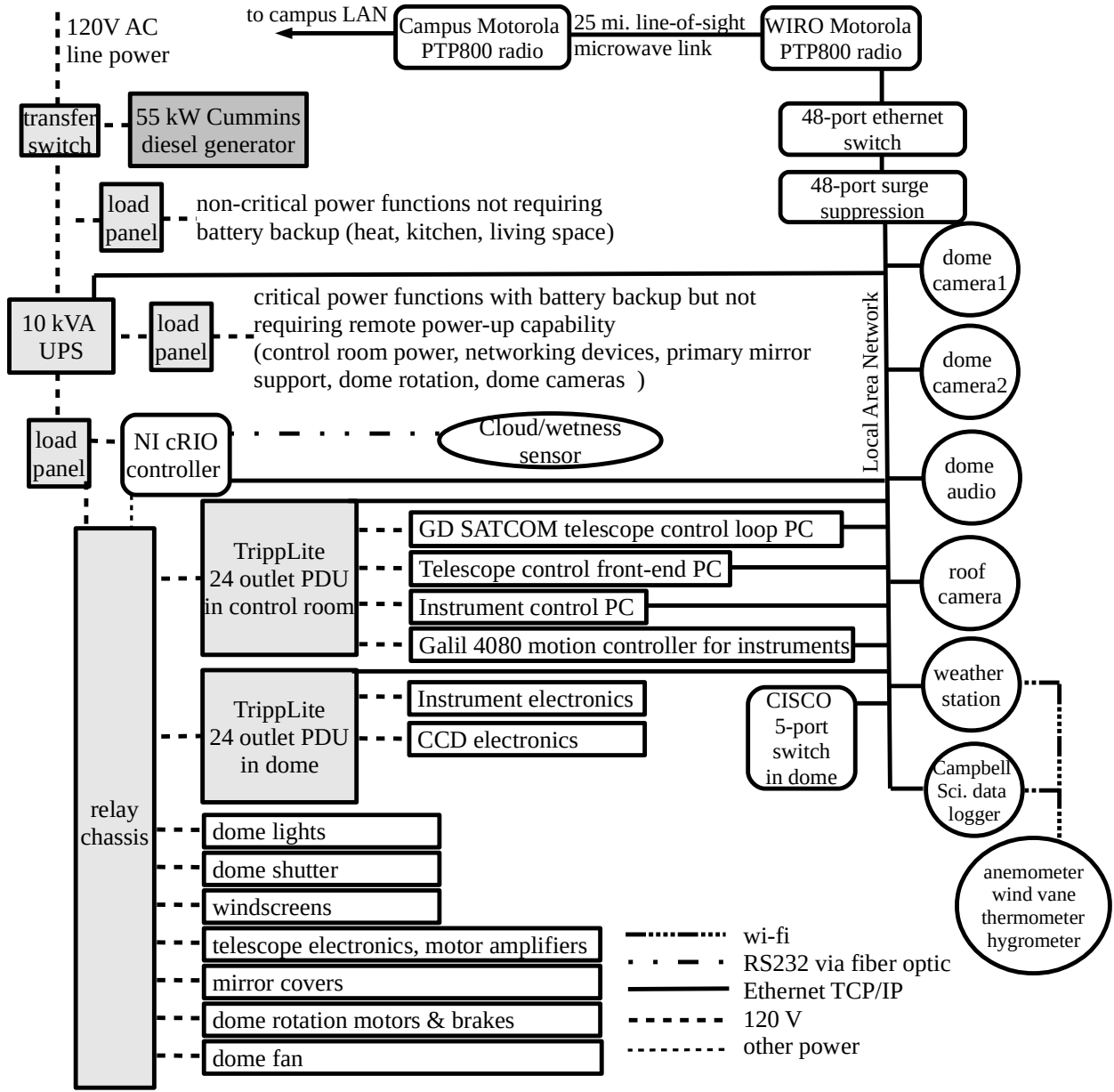}
\caption{A schematic overview of the upgraded power and control
systems that enable fully-remote observing at the WIRO telescope.  
Line styles given in the legend denote types of electrical power or 
communications protocols employed.    
\label{remoteschematic}}
\end{figure}

Figure~\ref{remoteschematic} provides a schematic of the
observatory power distribution and communications.  Heavy
dotted lines indicate 120 V power, thin dotted lines 
indicate control-voltage connections, and solid thin lines
show TCP/IP connections on the local area network (LAN).
Figure~\ref{remoteschematic} illustrates the layout of key
observatory infrastructure  but is not exhaustive; multiple
electrical circuits have been grouped into single logical
blocks for simplicity of presentation.   Power to each of
the key observatory devices is controlled by a bank of
Functional Devices, Inc.\ Relay-in-a-Box (RIB) relays
controlled using a National Instruments compactRIO 
(cRIO-9074) controller.  A custom LabVIEW graphical user
Observatory Control (OC) interface  written by the UW
Division of Research Support  allows a remote user to
monitor and control  each relay.   The OC GUI allows
activation of the dome shutters, mirror covers, dome lights,
dome fan, along with all  critical facility electrical
circuits.   The relays can also be controlled locally
through the use of a manual override switch that bypasses
the cRIO controller.   The OC GUI displays realtime status
on battery runtime (via TCP link to the UPS), weather (via
TCP link to the WIRO Davis Vantage Pro weather station), and
humidity/wetness (via an RS232 to optical fiber connection
that provides electrical isolation).   The OC GUI may be run
either locally on observatory computers or from the UW
campus remote observing computer.  An Aurora Eurotech Cloud
Sensor III cloud/wetness sensor  mounted on the roof
triggers (via the cRIO) a dome closure  if precipitation is
detected.  As an additional safety precaution, the telescope
control loop code includes a thread that monitors the network
connection to  the remote computer on campus;  it closes the
dome and stows the telescope in the event of a network failure.  

Lightning protection at the observatory is provided
site-wide by MCG Surge Protection devices and a
building-wide web of metal cable grounded to large copper
plates underneath the mountaintop. The local area network
ethernet switch and all ethernet devices are protected by an
AC Systems 48-port POE surge protection device that helps
eliminate damage from lightning-induced surges that have
historically occurred.  Additional surge suppression is
installed  on the RS232 and low-voltage analog signal
conductors leading from the dome motors and encoders to the
GD control system electronics; because of their length,
these are historically susceptible to lightning-induced
surges.

\subsection{Procedures for Fully Remote Operation}

\begin{figure}
\centering
\rotate
\includegraphics[width=7.in,angle=0]{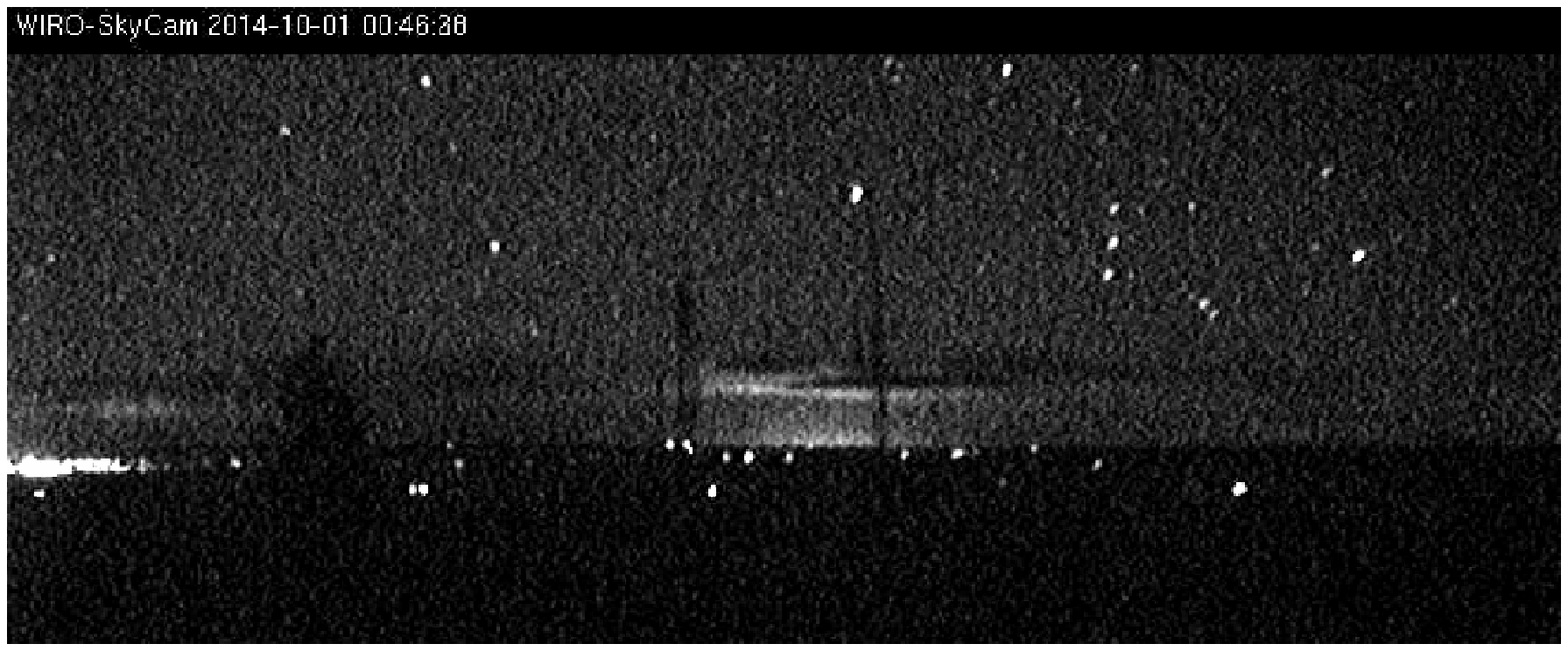}
\caption{An average of three 0.5 s exposures (slightly cropped in the 
horizontal and vertical directions) 
from the WIRO roof camera looking
east during new moon.  The lights of Laramie, WY, (far left) 
a thin cloud band illuminated by the lights of Cheyenne, WY (center),
and the rising constellations Gemini and Orion are visible.      
\label{night}}
\end{figure}

Remote operation requires that an observer on the UW campus
monitor observatory weather, observatory status,  and
instrument status in real time. The WIRO Davis 
Vantage Pro weather station provides real time temperature, 
humidity, wind speed and wind direction information to the 
OC GUI. An observer is required to close the dome if 
sustained wind speeds surpass 30 mph and the wetness sensor 
triggers dome closure if humidity exceeds $80\%$.
One AXIS Q6032-E network
camera\footnote{This product has been discontinued,
contact AXIS Communications to identify an equivalent.}
in a heated enclosure  mounted on the roof of the
living quarters allows a 360-degree view of the mountain
top, dome exterior, cloud conditions, and weather, even in
low-light conditions.  Figure~\ref{night} shows an average of three
0.5 s exposures from the roof camera facing east during New
Moon. The lights of Laramie, WY, are clearly visible at
far left, along with a thin cloud band illuminated by the
lights from Cheyenne, WY 60 miles to the east. The image
captures the
rising constellations Gemini and Orion.   The
image has been cropped by about 10\% and 30\%  in the
horizontal and vertical directions, respectively, but still serves to
illustrate the $>$2\degr\ field of view.     Two
additional AXIS Q6032-E network cameras are mounted on
opposite sides of the dome interior, allowing full
pan/tilt/zoom imaging of any portion of the telescope and
dome.   These cameras are used during startup and shutdown
sequences  to visually verify correct operation of mirror
covers, dome shutters, and other key mechanisms.  An AXIS
P8221 audio module inside the dome provides audio feedback
to a remote observer.

\section{Summary}

In the dawning era of time-domain astronomy,  WIRO is
positioned to make outsized returns as one of the largest
observatories operated by a single institution.

The ability
to implement flexible schedules or assign large blocks of
time (even hundreds of nights!) to a single program   will
allow the upgraded Wyoming Infrared Observatory to 
participate in campaigns of arbitrary cadence and duration. 
Future instrument and telescope upgrades will permit
visiting Cassegrain instruments a more favorable beam  once
a new wide-field f/9 secondary is constructed. As
historically has been the case, one of the strengths of the
Observatory will continue to  be student training. Graduate
and undergraduate students, along with visiting teams from
other institutions, will obtain hands-on experience in
telescope operation, programming, digital data analysis,
optics, instrument design, electronics, and real-time remote 
operation of a major scientific facility.  

\acknowledgements   This work was made possible by NASA 
through EPSCOR grant NNX10AM10H, and the National Science
Foundation through grants AST 0307778 and AST/PREST 0721281.
We are grateful to Charles Harmer for his assistance with
the  design of the prime-focus corrector.  We thank Bob Berrington
and Andy Monson for assistance with the installation
of the prime focus corrector.  We thank the Lander, WY
high school astronomy club, and their instructor, Joe Meyer, for 
inspiration to make WIRO a vehicle for learning and outreach
for students of all ages.  We thank Robert Gehrz, John Hackwell,
the State of Wyoming, and the National Science Foundation
for the vision to construct WIRO 40 years ago. Finally we
would like to thank the anonymous referee for their work in
helping to improve this paper.

\end{document}